\documentclass[conference]{IEEEtran}
\IEEEoverridecommandlockouts
\usepackage{cite}
\usepackage{amsmath,amssymb,amsfonts}
\usepackage{algorithmic}
\usepackage{graphicx}
\usepackage{textcomp}
\usepackage{xcolor}
\def\BibTeX{{\rm B\kern-.05em{\sc i\kern-.025em b}\kern-.08em
    T\kern-.1667em\lower.7ex\hbox{E}\kern-.125emX}}
\begin{document}

\title{Revisiting Disaggregated Large Language Model Serving for Performance and Energy Implications}

\author{\IEEEauthorblockN{Jiaxi Li}
\IEEEauthorblockA{UIUC \\
Urbana, Illinois \\
jiaxili3@illinois.edu}
\and
\IEEEauthorblockN{Yue Zhu}
\IEEEauthorblockA{IBM Research \\
Yorktown Heights, New York \\
Yue.Zhu@ibm.com}
\and
\IEEEauthorblockN{Eun Kyung Lee}
\IEEEauthorblockA{IBM Research \\
Yorktown Heights, New York \\
eunkyung.lee@us.ibm.com}
\and
\IEEEauthorblockN{Klara Nahrstedt}
\IEEEauthorblockA{UIUC \\
Urbana, Illinois \\
klara@illinois.edu}
}

\maketitle

\begin{abstract}
Different from traditional Large Language Model (LLM) serving that colocates the prefill and decode stages on the same GPU, disaggregated serving dedicates distinct GPUs to prefill and decode workload. Once the prefill GPU completes its task, the KV cache must be transferred to the decode GPU. While existing works have proposed various KV cache transfer paths across different memory and storage tiers, there remains a lack of systematic benchmarking that compares their performance and energy efficiency. Meanwhile, although optimization techniques such as KV cache reuse and frequency scaling have been utilized for disaggregated serving, their performance and energy implications have not been rigorously benchmarked. In this paper, we fill this research gap by re-evaluating prefill-decode disaggregation under different KV transfer mediums and optimization strategies. Specifically, we include a new colocated serving baseline and evaluate disaggregated setups under different KV cache transfer paths. Through GPU profiling using dynamic voltage and frequency scaling (DVFS), we identify and compare the performance–energy Pareto frontiers across all setups to evaluate the potential energy savings enabled by disaggregation. Our results show that performance benefits from prefill–decode disaggregation are not guaranteed and depend on the request load and KV transfer mediums. In addition, stage-wise independent frequency scaling enabled by disaggregation does not lead to energy saving due to inherently higher energy consumption of disaggregated serving.
\end{abstract}

% \begin{IEEEkeywords}
% TBD
% \end{IEEEkeywords}

\section{Introduction}
Large Language Models (LLMs) are transformer-based neural networks that process natural language inputs and generate coherent text outputs \cite{attention, GPT-4, DeepSeek-V3}. They are widely used in real-world applications such as conversation \cite{conversation_1, conversation_2, conversation_3} and code generation \cite{code_generation_1, code_generation_2, code_generation_3}. LLMs execute a two-stage inference process on GPUs to generate outputs. The compute-intensive prefill stage processes all input tokens in parallel and generates intermediate representations known as the key–value (KV) cache, while the memory-intensive decode stage generates output tokens autoregressively by repeatedly accessing the KV cache \cite{attention}. Two metrics are commonly used to evaluate LLM inference serving performance. Time to First Token (TTFT) evaluates prefill latency, and measures the latency between a request’s submission and the generation of the first output token, while Time per Output Token (TPOT) evaluates decode latency, and measures the average time to generate each subsequent token. LLM inference serving frameworks aim to minimize both TTFT and TPOT to ensure high Quality of Experience (QoE) for users.

LLM inference serving has been extensively studied for both performance and energy efficiency \cite{PagedAttention, SGLang, DistServe, EcoServe, DynamoLLM, GreenLLM-2}. Most existing frameworks \cite{PagedAttention, SGLang} colocate the prefill and decode stages on the same GPU to minimize KV cache transfer overhead. In contrast, recent studies \cite{DistServe, EcoServe, Mooncake, Splitwise, ShuffleInfer, semi-PD, TD-Pipe, TaiChi, WindServe, HeteroScale} advocate disaggregated serving, where individual GPUs are dedicated to either prefill or decode workloads. This design offers two primary benefits: (1) it eliminates the interference that occurs when prefill and decode stages share the same GPU, and (2) it enables independent configuration of prefill and decode GPUs to better satisfy their respective Service Level Objectives (SLOs) on serving performance. However, disaggregation also introduces unavoidable overhead, as the KV cache must be transferred from the prefill GPU to the decode GPU for each request.

Existing benchmarking studies \cite{DistServe, EcoServe} report that disaggregating prefill and decode into separate GPUs leads to improved LLM serving performance compared to colocated execution. However, these studies have explored only a limited set of experimental setups, while still missing performance analysis under certain scenarios. First, most existing works only utilize high-speed links (e.g., NVLink) to transfer KV cache from the prefill GPU to the decode GPU \cite{DistServe, Mooncake, ShuffleInfer, semi-PD, TaiChi, WindServe}. Although industry has provided storage solutions that support KV cache transfer via CPU memory offloading \cite{Mooncake, CacheBlend, CachedAttention, PromptCache, Pensieve, RAGCache, Wild}, there remains a lack of benchmarking studies that compare the performance and energy efficiency between different transfer paths. Such analysis is essential for guiding LLM serving strategies on hardware platforms that lack high-speed GPU peer-to-peer interconnects. Second, existing works have tested limited cases of colocated serving baselines. Specifically, using a single NVIDIA A100 GPU, DistServe \cite{DistServe} reported improved TTFT performance in the prefill-only case than in the colocated case, and similarly better TPOT in the decode-only case. However, in DistServe’s setup, disaggregated serving requires at least two GPUs, one for prefill and one for decode, to realize the claimed latency improvements, whereas the colocated baseline used only a single GPU. Another colocated serving baseline with two GPUs should also be included in the benchmarking study.
 
% Meanwhile, existing studies lack a comprehensive benchmarking analysis that evaluates the performance and energy efficiency benefits of KV cache reuse. KV caches computed from previous requests can be stored and reused for future requests with similar contexts, thereby avoiding redundant prefill computations and potentially improving serving performance \cite{CacheBlend, CacheGen}. Many recent LLM serving frameworks \cite{Mooncake} already support such reuse through dedicated KV cache stores in CPU memory or disk. Compared to high-speed interconnects like NVLink, CPU memory and disk provide significantly larger capacity to enable more aggressive KV cache reuse. However, the two-way offloading involved in these approaches also introduces additional transfer latency. As a result, a trade-off analysis is crucial to determine under what conditions offloading-based KV cache reuse does yield performance and energy gains. However, that remains unexplored in existing literature.
% [M]: KV reuse paragraph commented and used for full conference submission

Moreover, while Dynamic Voltage and Frequency Scaling (DVFS) has proven effective for improving energy efficiency in colocated LLM serving, its potential in the disaggregated setting remains largely unexplored in existing works. Prior works \cite{DynamoLLM, BoFL, throttLL’eM, SLO-Aware, Greener-LLMs} have utilized DVFS to profile the inference latency and energy consumption at different GPU frequencies, based on which they constructed the latency-energy Pareto frontier. As illustrated in Figure \ref{fig:exp2-latency}, these frontiers typically exhibit a U-shaped curve reflecting the trade-off between latency performance and energy consumption. During inference, GPU frequency is set according to the frontier to meet all latency SLOs at the minimal energy usage. However, this approach falls short on processing requests with imbalanced latency SLOs (e.g., tight TTFT but relaxed TPOT). Meeting the TTFT requirement often forces the GPU to run at a higher frequency, which leads to unnecessary energy consumption from overprovisioned TPOT performance. Disaggregated serving could potentially mitigate this inefficiency through independent frequency optimization. By assigning separate GPUs to prefill and decode, each GPU is allowed to operate at a different frequency that satisfies its own SLO with minimal energy cost.

To fill the research gap, we conduct extensive benchmarking studies to re-evaluate prefill–decode disaggregation under various experiment setups and optimization strategies. First, we include a new colocated serving baseline that employs two GPUs, with each processing half of the requests in the batch. The baselines are compared against disaggregated serving setups under different KV transfer paths (Section \ref{sec:varying-batch-size}). Furthermore, we assess the benefits of independent frequency optimization through DVFS. By conducting offline profiling at multiple GPU frequencies, we construct latency–energy Pareto frontiers for both prefill and decode stages. We iterate this for each setup, and compare the results across colocated and disaggregated serving to understand their energy trade-offs (Section \ref{sec:varying-gpu-frequency}).
% [M]: Write experiment description of the KV reuse experiment

Our key takeaways from the benchmarking experiments are summarized as follows.

\begin{itemize}
    \item \textbf{The benefits of prefill–decode disaggregation are not universal and depend on the request load and KV transfer medium.} Disaggregation doubles the request load per GPU, resulting in degraded TTFT performance when high-speed interconnects for KV cache transfer are unavailable. However, as number of requests per batch increases, colocated serving suffers from KV cache eviction and recomputation due to limited GPU memory, resulting in degraded TPOT performance.
    \item \textbf{Independent frequency scaling does not necessarily lead to energy savings.} Even with stage-wise frequency scaling, disaggregated serving could still consume more time in the prefill stage than the colocated baseline, resulting in higher overall energy consumption.
\end{itemize}

\section{Background}

\subsection{LLM Inference}
Large Language Models (LLMs) are transformer-based neural networks with a vast number of parameters, which are capable of handling sophisticated tasks such as conversation \cite{conversation_1, conversation_2, conversation_3}, code generation \cite{code_generation_1, code_generation_2, code_generation_3}, and summarization \cite{summarization_1, summarization_2}. To respond to user requests, LLMs undergo an inference process that can be divided into two stages. In the prefill stage, the entire input token sequence is processed in parallel, producing the key–value (KV) cache as the intermediate representation. KV cache contains the key (K) and value (V) tensors generated at each attention layer, which encode the contextual information of the input. In the subsequent decode stage, the model generates output tokens one by one in an autoregressive manner by reusing the stored KV cache.

The computational characteristics of the two stages differ significantly. The prefill stage is compute-heavy, and is dominated by large matrix multiplications across the full sequence of input tokens. By contrast, the decode stage is memory-heavy, as each new token generation requires repeatedly looking up the entire KV cache, which must therefore remain in GPU memory throughout the process.

To support large-scale deployment, LLMs are served through specialized inference systems that manage scheduling and resource allocation. Their performance and scalability are commonly evaluated using two metrics. Time to First Token (TTFT) is a metric for the prefill stage, which measures the latency from when a request is submitted until the first output token is generated. TTFT reflects the responsiveness of a serving system and is especially important for real-time interactive applications. Time per Output Token (TPOT) evaluates the decode stage, which measures the average time between consecutive output tokens. When TPOT matches or exceeds human reading speed, responses feel smooth and natural, whereas higher TPOT causes pauses that disrupt readability. To ensure acceptable Quality of Experience (QoE), requests may specify SLOs that require the serving system to meet certain latency or throughput targets.

\subsection{Prefill and Decode Disaggregation}\label{sec:p_and_d_disagg}
Conventional LLM inference serving systems colocate the prefill and decode stages on the same GPU or GPU node \cite{PagedAttention, SGLang}. Recent work, however, has proposed disaggregated serving, in which each GPU worker is dedicated exclusively to either prefill or decode. Studies \cite{DistServe, EcoServe} have demonstrated several benefits of this design. First, disaggregation eliminates prefill–decode interference. For example, \cite{DistServe} observed significant degradation in decode latency when a single prefill request was introduced into a batch of decode workloads, and vice versa. By separating these stages onto different GPU workers, prefill–decode interference is avoided and each device can be fully optimized for its respective workload. Second, disaggregation enables flexible GPU optimization. As requests may impose distinct SLOs for prefill and decode latency, GPU workers can be configured differently—e.g., with stage-specific batching strategies \cite{EcoServe}, parallelism levels \cite{DistServe}, or frequency scaling policies \cite{DynamoLLM}—to maximize performance for each stage.

However, prefill-decode disaggregation also introduces a new overhead, as the KV cache produced during prefill must be transferred to the decode worker. For recent billion-parameter LLM models, the KV cache can be very large, and transmitting it across GPUs may lead to significant latency increase. To address this, two classes of KV transfer strategies have emerged, depending on the hardware topology of the cluster. In multi-node clusters where prefill and decode are placed on different GPU nodes, KV transfer relies on high-speed interconnects such as RDMA \cite{Mooncake, HeteroScale} or InfiniBand \cite{Splitwise}. In single-node clusters where prefill and decode are disaggregated across GPUs within the same node, KV transfer is mainly realized through GPU peer-to-peer links such as NVLink \cite{DistServe, Mooncake, ShuffleInfer, semi-PD, TD-Pipe, TaiChi, WindServe} or PCIe \cite{EcoServe, ShuffleInfer, TD-Pipe, WindServe}.

\subsection{KV Cache Sharing and Reusing}\label{sec:kvc_sharing_and_reusing}
Beyond prefill–decode disaggregation, the computation overhead of LLM inference can be further optimized through KV cache sharing and reuse. The KV cache, generated during the prefill stage, stores intermediate context information and can be reused across requests with overlapping input tokens to avoid redundant computation. This technique is particularly effective for workloads with long shared contexts. For example, in retrieval-augmented generation (RAG) scenario, a user prompt is augmented with retrieved documents (e.g., legal texts, news articles, or Wikipedia pages). Although the prompts may differ across users, the retrieved documents are often identical, allowing their KV caches to be reused rather than recomputed.

Existing research has mainly explored two strategies for KV cache reuse. The simplest form is prefix matching: when the beginning of a new request exactly matches the beginning of a prior request, the cached KV tensors for those prefix tokens can be reused to avoid redundant computation \cite{CachedAttention, SGLang, RAGCache, Wild}. While straightforward, prefix matching often yields limited benefits, as user prompts may diverge in their opening tokens even if they later converge on overlapping content. To overcome this limitation, more recent work has proposed Position-Independent Caching (PIC), which allows reuse of KV cache for matching tokens regardless of their position in the input sequence \cite{CacheBlend, PromptCache, Pensieve, EPIC}. This is often achieved by selectively recomputing cross-attention between reused KV blocks and surrounding tokens \cite{CacheBlend}.

\subsection{Energy-efficient LLM Inference}
Unlike the one-time cost of training, LLM inference dominates resource consumption because it requires continuous GPU computation to serve millions of requests per minute. At datacenter scale, this translates into significant energy demand. For instance, a modern datacenter with 3,000 NVIDIA Blackwell B200 GPUs can draw approximately 86,400 kWh per day \cite{B200_Energy}, an amount comparable to the daily electricity usage of a small U.S. town with about 2,800 households \cite{US_Energy}. Such energy demands impose heavy financial burdens on AI companies \cite{Company_Energy_Cost_1}, while also raising environmental concerns due to the associated carbon footprint \cite{Environment_Energy_Cost_1}. Consequently, energy-efficient LLM inference has become an active research focus in both industry and academia.

Recent efforts to improve energy efficiency fall into two complementary categories. Workload-optimization approaches \cite{GreenLLM, Sustainable-LLM, StoreLLM, EmbAdvisor, GreenLLM-2} aim to reduce the amount of computation required per request while preserving the SLOs on accuracy. Techniques used include layer dropping \cite{GreenLLM}, model quantization \cite{Sustainable-LLM}, and reuse of previously computed results \cite{StoreLLM, EmbAdvisor}. These approaches directly shrink the per-request workload, thereby lowering both compute overhead and energy usage. System-optimization approaches \cite{DynamoLLM, throttLL’eM, SLO-Aware, Greener-LLMs, Offline, Hybrid-Heterogeneous} instead focus on adjusting the serving infrastructure. These works tune GPU frequency \cite{DynamoLLM, throttLL’eM, SLO-Aware, Greener-LLMs}, GPU parallelism level \cite{DynamoLLM, Greener-LLMs}, batching strategy \cite{Greener-LLMs}, and request routing strategy \cite{Offline, Hybrid-Heterogeneous} to maximize power efficiency under latency constraints. Rather than reducing the workload itself, system-optimization approaches reduce the energy cost per unit of work performed.

\section{Related Work and Our Motivation}
Based on benchmarking experiments, existing works report performance gains and energy efficiency from prefill–decode disaggregation compared to colocation \cite{DistServe, EcoServe, GreenLLM-2}. In this section, however, we argue that existing studies have based their benchmarking experiments on specific baseline assumptions, which limits the generalizability of their results across different deployment scenarios. Meanwhile, while existing studies have supported disaggregated LLM serving with various KV transfer paths and optimization techniques such as KV cache reuse and frequency scaling, each approach presents distinct trade-offs. A comprehensive benchmarking study is therefore necessary to systematically evaluate these trade-offs and identify the most effective serving configurations for different deployment scenarios. However, such an analysis is still lacking in existing studies.

\subsection{Unexplored Baseline Configurations}
Existing works have compared disaggregated serving to a limited case of colocated serving baseline. DistServe \cite{DistServe} represents a state-of-the-art disaggregated LLM serving framework. In its benchmark study, a single NVIDIA A100 GPU was used to serve a 13B-parameter model under a synthetic workload. Three cases were evaluated for the GPU: prefill-only, decode-only, and colocation. The reported results suggested that dedicating the GPU exclusively to prefill or decode improved TTFT or TPOT compared to the colocated case. Although their results are reasonable, we argue that they are missing one important colocated serving baseline. In their experiment setting, achieving the claimed performance gains from prefill–decode disaggregation requires at least two GPUs—one for prefill and one for decode. By contrast, the original colocated baseline used only a single GPU. To fairly assess disaggregation and colocation under equivalent GPU resources, we propose to include a new colocated baseline in which two GPUs each perform both prefill and decode, with the incoming requests evenly distributed between them. Section \ref{sec:varying-batch-size} and \ref{sec:varying-gpu-frequency} presents a comprehensive evaluation of this new baseline relative to disaggregated alternatives.

\subsection{Benchmarking KV Cache Transfer Paths}
As discussed in Section \ref{sec:p_and_d_disagg}, existing intra-node disaggregated serving frameworks \cite{DistServe, Mooncake, ShuffleInfer, semi-PD, TaiChi, WindServe} primarily rely on high-speed GPU peer-to-peer links for KV cache transfer. While effective, this approach is constrained by hardware availability, as interconnects such as NVLink are not present on all commodity GPUs \cite{TD-Pipe}. On the other hand, CPU memory and disk are resources that are widely available in modern compute nodes. Based on that, industry has proposed offloading-based solutions to support disaggregated serving. For example, Mooncake \cite{Mooncake} supports prefill-to-decode KV cache transfer via both NVLink and CPU memory offloading. However, to the best of our knowledge, there is no prior work that focuses on benchmarking and comparing the perfofrmance and energy efficiency across different KV transfer paths inside disaggregated serving. We therefore conduct a comprehensive experiment to fill this research gap in Section \ref{sec:varying-batch-size}. Our findings are particularly relevant to GPU cluster operators, as they highlight the trade-off between performance and cost. While NVLink can deliver superior latency performance, its high hardware cost and potential elevated energy consumption may ultimately result in greater overall operating expenses.

\subsection{Benchmarking Disaggregated Frequency Scaling}
DVFS has been extensively explored for prefill–decode colocated serving and has proven effective in reducing energy consumption \cite{DynamoLLM, throttLL’eM, SLO-Aware, Greener-LLMs}. However, its effectiveness in disaggregated LLM serving remains largely unexplored.

DVFS is a hardware-level power-management technique that adjusts a processor’s operating frequency and supply voltage at runtime \cite{BoFL}. By lowering frequency and voltage, DVFS reduces power consumption at the cost of lower performance on the workload (e.g., higher TTFT or TPOT). Existing works take this opportunity to design energy-efficient SLO-aware LLM inference systems \cite{DynamoLLM, throttLL’eM, SLO-Aware, Greener-LLMs}. In these systems, offline profiling is first used to construct the performance–power Pareto frontiers by measuring serving performance and power consumption across different GPU frequencies. During online serving, the GPU frequency is then selected along the frontiers to satisfy performance SLOs while minimizing energy consumption. However, these approaches may exhibit performance degradation when serving requests with imbalanced SLO requirements between prefill and decode. In real-world scenarios, prefill and decode often operate under different latency constraints. For instance, reducing inference response time (TTFT) always improves user experience, but increasing token generation speed (TPOT) beyond human reading or listening speed provides little additional benefit \cite{Andes}. This creates cases where requests demand tight TTFT but only moderate TPOT requirements. In colocated serving, however, GPU frequency scaling affects both stages simultaneously. As a result, the GPU is often configured at a high frequency to meet TTFT constraints, which unnecessarily accelerates TPOT and causes excessive energy consumption due to overprovisioned performance.
% [M]: One example figure of a performance–power Pareto frontier]

Prefill–decode disaggregation could potentially address this limitation, as it naturally enables independent frequency optimization. Different frequencies can be assigned to the prefill and decode GPUs so that each stage meets its respective SLO with minimal power consumption. Without violating latency requirements, prefill–decode disaggregation provides the opportunity for more aggressive energy savings. To evaluate this potential, we conduct a comprehensive benchmarking study to measure the performance–power Pareto frontiers of disaggregated serving and compare them against those of colocated serving (Section \ref{sec:varying-gpu-frequency}).
% [M]: Better description for experiment that benchmarks disaggregated frequency scaling]

\subsection{Motivation Summary}
In summary, prior studies have provided valuable insights on prefill-decode disaggregation but remain incomplete due to unexplored baseline configurations and the absence of a comprehensive benchmarking study that examines the trade-offs among different KV cache transfer paths and optimization techniques. To address this research gap and develop a deeper understanding of the performance and energy implications of prefill–decode disaggregation, we conduct a series of benchmarking experiments aimed at identifying when and which KV transfer paths or optimization techniques are most effective under different scenarios. The details of these experiments are presented in the following sections.

\section{Benchmark Methodology}

\subsection{Overview}
Our benchmarking study consists of two main experiments:

\textbf{Experiment 1 (Section \ref{sec:varying-batch-size}):}
We benchmark the performance and energy consumption of disaggregated LLM serving under different experimental setups with each differ in KV cache transfer paths, and compare the results with the colocated baselines. We evaluate each experimental setup with increasing request batch sizes that lead to KV caches of varying sizes, and record the resulting performance and energy consumption for each case. Based on that we analyze how the benefits of prefill–decode disaggregation evolve as KV cache size grows.

\textbf{Experiment 2 (Section \ref{sec:varying-gpu-frequency}):}
We further evaluate the impact of GPU frequency scaling on performance and energy efficiency between disaggregation and colocation. For each experimental setup, we fix the workload and measure latency and energy consumption at multiple GPU frequency settings. From these measurements, we construct latency–energy Pareto frontiers for both prefill and decode stages. We then compare these frontiers across experimental setups to assess the potential energy-saving benefits in disaggregated serving that are enabled by independent frequency scaling.

\subsection{Hardware}
The benchmark experiment is conducted on a cluster node with two NVIDIA A100 GPUs, each with a GPU memory of 40 GB. These two GPUs are connected to the same PCIe bridge that supports PCIe Gen3. These GPUs support a wide range of frequency scaling from 0.21 GHz to 1.41 GHz. The cluster is also equipped with a 18-cores Intel Xeon E5-2697 v4 CPU with base frequency at 2.3 GHz and boost frequency at 3.6 GHz, as well as a Samsung 1 TB DDR4 Memory and a Samsung 3.2 TB NVMe SSD.

\subsection{Software Stack}
\textbf{vLLM \cite{vLLM}:} vLLM is an open-source LLM inference engine widely used in both academia and industry. It introduces PagedAttention \cite{PagedAttention}, an optimized memory management system that virtualizes the KV cache storage and organizes it in fixed-size pages to minimize memory fragmentation. It also supports continuous batching to efficiently handle variable sequence lengths. In our experiments, each GPU serves a single vLLM process responsible for executing the main inference pipeline. We further adapt vLLM’s benchmarking utility to generate synthetic datasets and collect performance metrics.

\textbf{LMCache \cite{LMCache}:} LMCache is an open-source KV cache management engine. It introduces CacheGen \cite{CacheGen} for efficient KV cache compression as well as CacheBlend \cite{CacheBlend} to support position-independent reuse for KV cache.  LMCache also provides vLLM-compatible connectors that enable KV cache transfer between vLLM processes. In our experiments, we adapt these connectors to implement different KV cache transfer paths inside disaggregated LLM serving.

\subsection{Model and Dataset}
Llama \cite{LLaMA} is a widely used family of open-source large language models released by Meta AI. In our experiments, we specifically employ Llama-3.2-3B \cite{Llama-3.2-3B} as the target model for benchmarking. To generate controlled and reproducible workloads, we use synthetic data produced by RandomDataset, a dataset generator released from the vLLM stack, which generates synthetic requests by sampling random token sequences of configurable batch size, input length, and output length. In our setup, we use RandomDataset to generate inference workloads with varying batch sizes, and we set the request rate to infinite so that all requests are dispatched to the GPU worker simultaneously.

\subsection{Metrics}
We use Time to First Token (\textbf{TTFT}) to evaluate prefill latency and Time per Output Token (\textbf{TPOT}) to evaluate decode latency. We also report \textbf{Prefill Throughput} and \textbf{Decode Throughput} respectively. \textbf{Energy Consumption} is obtained by integrating instantaneous power readings over the inference period. We report energy usage across different hardware components. Specifically, instantaneous GPU power is measured by pynvml \cite{pynvml}, a Python library that provides access to NVIDIA’s Management Library (NVML) \cite{NVML}. Instantaneous power by CPU and DRAM is measured by reading the Running Average Power Limit (RAPL) interface \cite{RAPL} that is supported by Intel processors. We also record the instantaneous total power of the cluster node via reading the Intelligent Platform Management Interface (IPMI) \cite{IPMI}. For each inference workload, we report energy consumption in joules per token, calculated by dividing the total energy consumed by the total number of tokens processed, including both input and output tokens.

\subsection{Experimental Configurations Setup}
% \textbf{\textit{co-1gpu}} represents the prefill-decode colocated serving baseline used in existing works \cite{DistServe}. This baseline utilizes one GPU to serve a vLLM V1 inference engine \cite{vLLM} that performs both prefill and decode locally. In this configuration, the GPU allocates 28 GB of memory for KV cache storage.
% [M]: Uncomment in the final conference version

\textbf{\textit{co-2gpus}} represents our new colocated serving baseline. Different from \textit{co-1gpu}, this baseline utilizes two GPUs, which provides equivalent GPU resources to that of disaggregated serving setups. In this baseline, each GPU serves a vLLM V1 inference engine that performs both prefill and decode locally. With two complete inference processes, the total batch is evenly divided between the two GPUs, with each handling half of the incoming requests to reduce their compute and memory overhead. In this configuration, each GPU allocates 28 GB of memory for KV cache storage.

\textbf{\textit{dis-gpu}} represents the disaggregated serving setup that utilizes GPU peer-to-peer (P2P) communication for KV cache transfer. Specifically, one GPU is dedicated for prefill and the other for decode. P2P KV cache transfer is enabled through a shared PCIe bridge connecting both GPUs within the same node. The software pipeline is implemented using vLLM V1 \cite{vLLM} and LMCacheConnectorV1 \cite{LMCache} that orchestrates the inter-GPU communication flow. Underneath, it leverages the NVIDIA Inference Xfer Library (NIXL) \cite{NIXL} to perform efficient data transfer via the CUDA Inter-Process Communication (cuda\_ipc) \cite{cuda_ipc} transport. This integration allows GPUs to directly access each other’s device memory to achieve higher transfer bandwidth for KV cache exchange.

\textbf{\textit{dis-cpu}} represents the disaggregated serving setup that realizes KV cache transfer through CPU offloading. Similar to \textit{dis-gpu}, one GPU is for prefill and the other is for decode. After the prefill stage, the KV caches are offloaded to the CPU DRAM. The decode GPU loads them back when it starts generating the output tokens. We utilize vLLM V1 and LMCacheConnectorV1 to orchestrate the software pipeline. The allowable CPU memory used for KV cache storage is set to 125 GB. In addition, a shared Redis \cite{Redis} lookup server is deployed between the GPUs, where the prefill GPU updates the lookup table after offloading a KV cache block to CPU DRAM, and the decode GPU queries the table to locate and fetch the required KV cache blocks.

\textbf{\textit{dis-disk}} represents the disaggregated serving setup that transfers KV cache through disk offloading. Similar to other disaggregated baseline, one GPU is responsible for prefill and the other is for decode. After the prefill stage, KV caches are first offloaded to the disk, and then offloaded back to the decode GPU. For a fair comparison to \textit{dis-cpu}, we force the kernel to bypass the page cache and perform the complete disk access each time when reading or writing KV caches. vLLM V1 and LMCacheConnectorV1 are employed to implement software-level KV cache transfer, with a maximum of 125 GB of disk space allocated for cache storage. The fs\_connector \cite{vLLM} is used to coordinate KV cache storage and retrieval. Acting as middleware running in CPU DRAM, fs\_connector receives KV caches from the prefill GPU before writing them to disk and fetches the corresponding caches from disk to offload them back to the decode GPU.

\section{Benchmark Results}

\subsection{Experiment 1: Varying Batch Size}\label{sec:varying-batch-size}

\begin{figure}
    \center{
        \includegraphics[width=250pt]{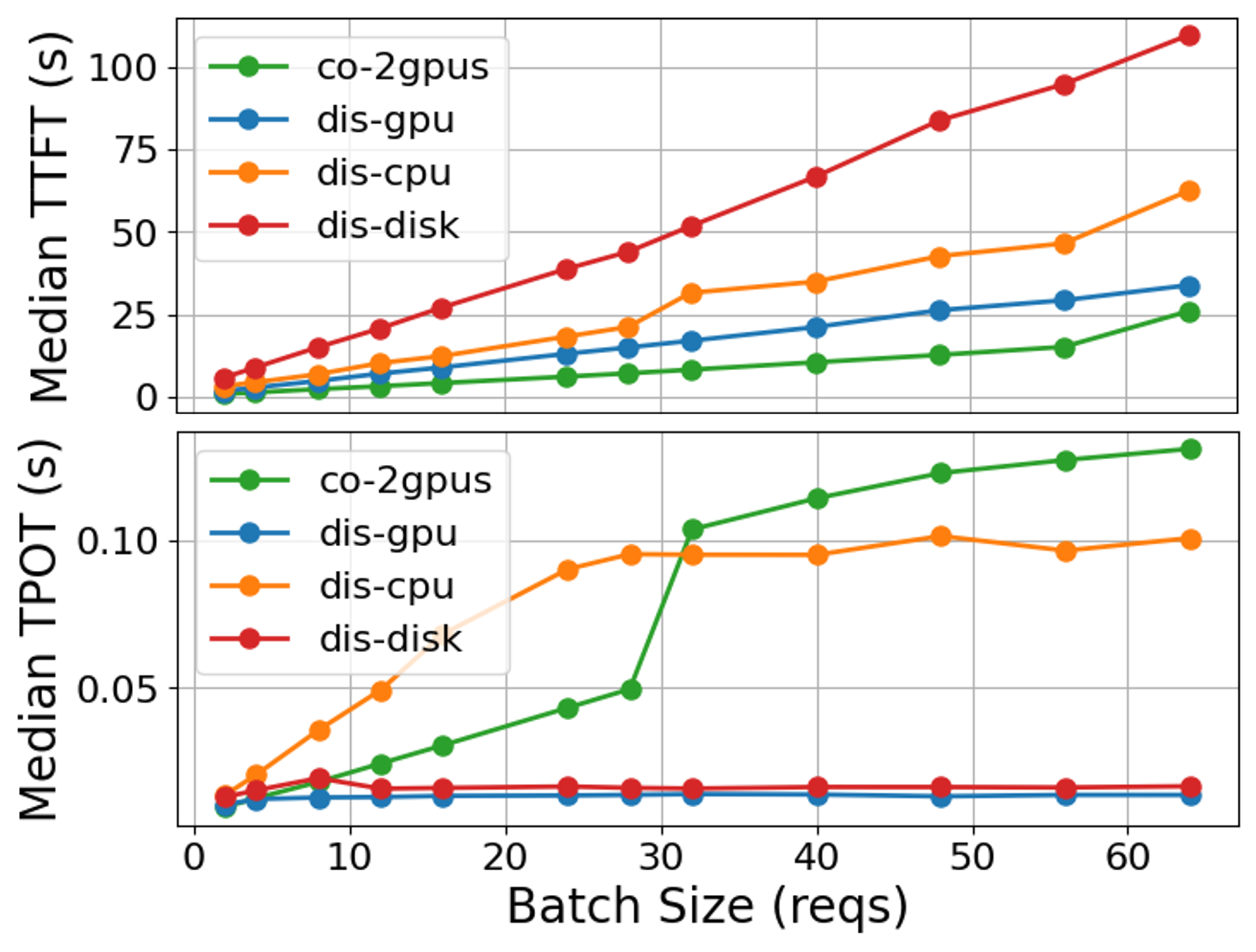}
    }
    \caption{TTFT and TPOT performance as batch size increases across different experimental setups.}
    \label{fig:exp1-latency}
\end{figure}

\begin{figure}
    \center{
        \includegraphics[width=250pt]{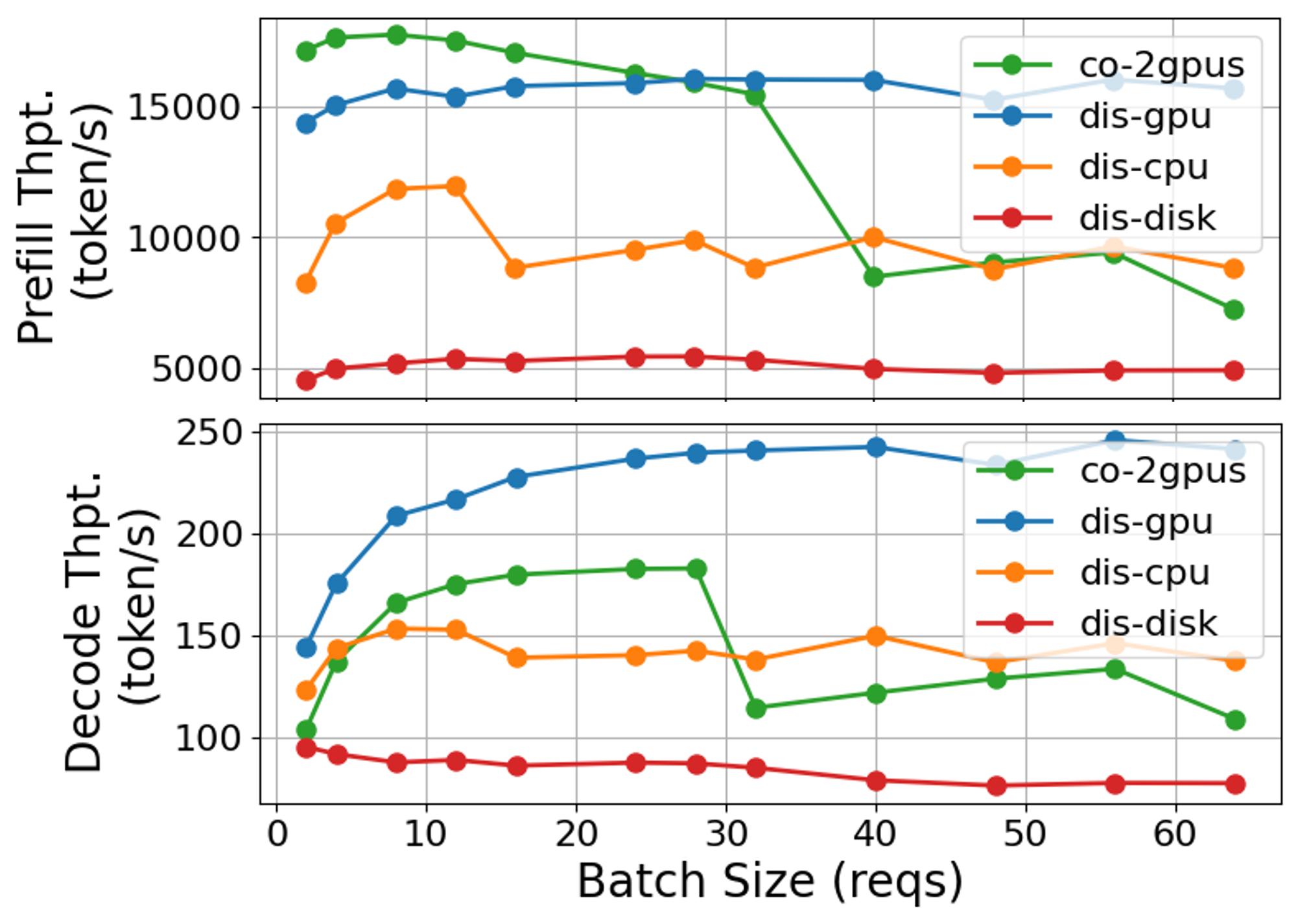}
    }
    \caption{Prefill and decode throughput as batch size increases across different experimental setups.}
    \label{fig:exp1-throughput}
\end{figure}

\begin{figure}
    \center{
        \includegraphics[width=250pt]{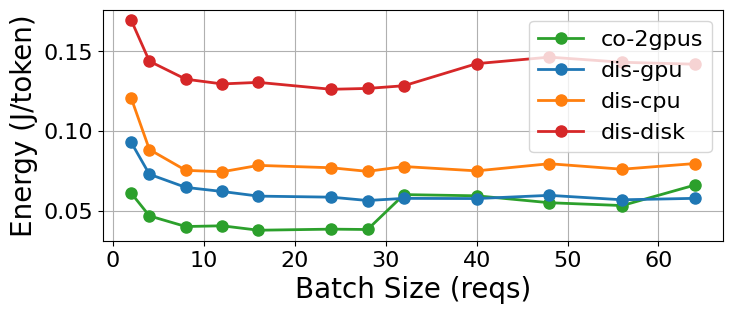}
    }
    \caption{Total energy consumption as batch size increases across different experimental setups.}
    \label{fig:exp1-energy}
\end{figure}

\begin{figure}
    \center{
        \includegraphics[width=250pt]{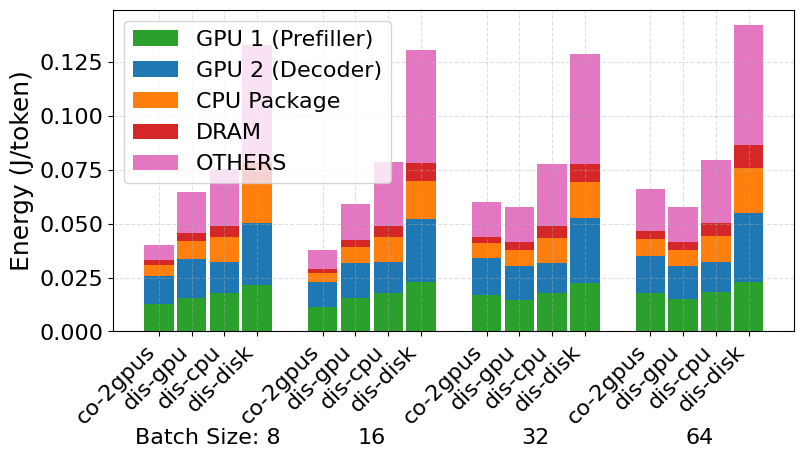}
    }
    \caption{Energy contributed by each hardware component as batch size increases across different experimental setups.}
    \label{fig:exp1-breakdown}
\end{figure}

We present and analyze the results of Experiment 1, which evaluates the performance and energy consumption of different experimental setups under increasing batch sizes.

Figure \ref{fig:exp1-latency} represents the latency result, with the upper sub-figure showing TTFT in the y-axis and the lower sub-figure showing TPOT in the y-axis. Each request is fixed to include an input token size of 16,384 and an output token size of 256. To investigate how KV cache size affects the serving performance and energy efficiency, we increase the batch size (represented by the x-axis) from 2 to 64, leading to the total size of generated KV cache ranging from 3.5 GB to 112 GB. Each line indicates the latency performance achieved by a specific setup, as indicated in the figure legend.

We first observe that for all setups, median TTFT increases as batch size increases. This is expected, as with more concurrent requests sent to the prefill GPU, less computation resources are allocated to each request, leading to higher latency. Also, among the disaggregated setups, \textit{dis-gpu} achieves the best TTFT because it only passes one PCIe bridge without touching deeper memory tiers. After that comes the CPU offloading based setup (\textit{dis-cpu}), and then the disk offloading based setup (\textit{dis-disk}). To our surprise, for all cases of batch sizes, \textit{co-2gpus} is achieving the best TTFT performance, which indicates that prefill-decode disaggregation does not always guarantee performance gains. Under equivalent GPU resources, each GPU under the \textit{co-2gpus} baseline only needs to process half a batch of the requests. Even with prefill-decode interference, the reduced batch size significantly reduces computation overhead, ultimately leading to an improved TTFT compared to disaggregated setups.

For TPOT, we notice that only \textit{co-2gpus} shows a sharp increase when batch size exceeds 32. This is because the total size of the generated KV cache is larger than the allocated GPU memory, causing the previously generated KV cache to be evicted. This part of the KV cache has to be recomputed when the decode stage later requires them, causing a significant delay in token generation. In addition, we observe that \textit{dis-disk} exhibits a TPOT trend similar to \textit{dis-gpu} and performs significantly faster than \textit{dis-cpu}. This result is unexpected, as disk offloading places the KV cache farther from GPU memory and would typically incur higher latency. This phenomenon is currently under investigation.

Summarizing the latency results leads to our first takeaway: performance gains from prefill-decode disaggregation highly depends on batch size. Under equivalent GPU resources, disaggregated serving could double the batch size compared to colocated serving, resulting in higher overall TTFT. On the other hand, colocated serving suffers performance degradation at larger batch sizes when the amount of KV cache exceeds allocated GPU memory capacity.

Figure \ref{fig:exp1-throughput} shows the throughput result, with the upper sub-figure for prefill and lower sub-figure for decode. We first observe that, across all disaggregated setups, throughput increases with batch size up to 16, beyond which it plateaus as the GPU’s computational capacity becomes saturated. However, due to the aforementioned GPU memory limitation, \textit{co-2gpus} experiences a significant drop in throughput when the batch size reaches around 32. In addition, similar to latency, \textit{dis-gpu} achieves the best performance among disaggregated setups because of the shortest transfer path, followed by \textit{dis-cpu} and\textit{dis-disk}.

Figure \ref{fig:exp1-energy} shows the total energy consumed by different experimental setups. Across all setups, energy consumption initially decreases and then converges as batch size increases. This behavior arises because static, or standby, energy is amortized over a larger amount of computation as the GPU becomes more fully utilized during inference. For \textit{co-2gpus}, we also observe a sharp increase in energy consumption when batch size reaches 32, due to KV cache eviction and recomputation. Among disaggregated setups, \textit{dis-gpu} again achieves the best performance because it takes the least latency to complete the inference workload, and is followed by \textit{dis-cpu} and\textit{dis-disk}.

To further analyze the energy consumption contributed by each hardware component, we present the energy consumption breakdown in Figure \ref{fig:exp1-breakdown}. Each bar in the figure represents the energy consumption of an experimental setup under a batch size. Inside each bar, boxes of different colors indicate the energy consumption contributed by each hardware component, as described in the figure legend. We observe that, as the KV cache transfer path touches deeper memory tiers, it consumes more energy from non-GPU hardwares, including CPU, DRAM, and disk.

\subsection{Experiment 2: Varying GPU Frequency}\label{sec:varying-gpu-frequency}

\begin{figure*}
    \center{
        \includegraphics[width=500pt]{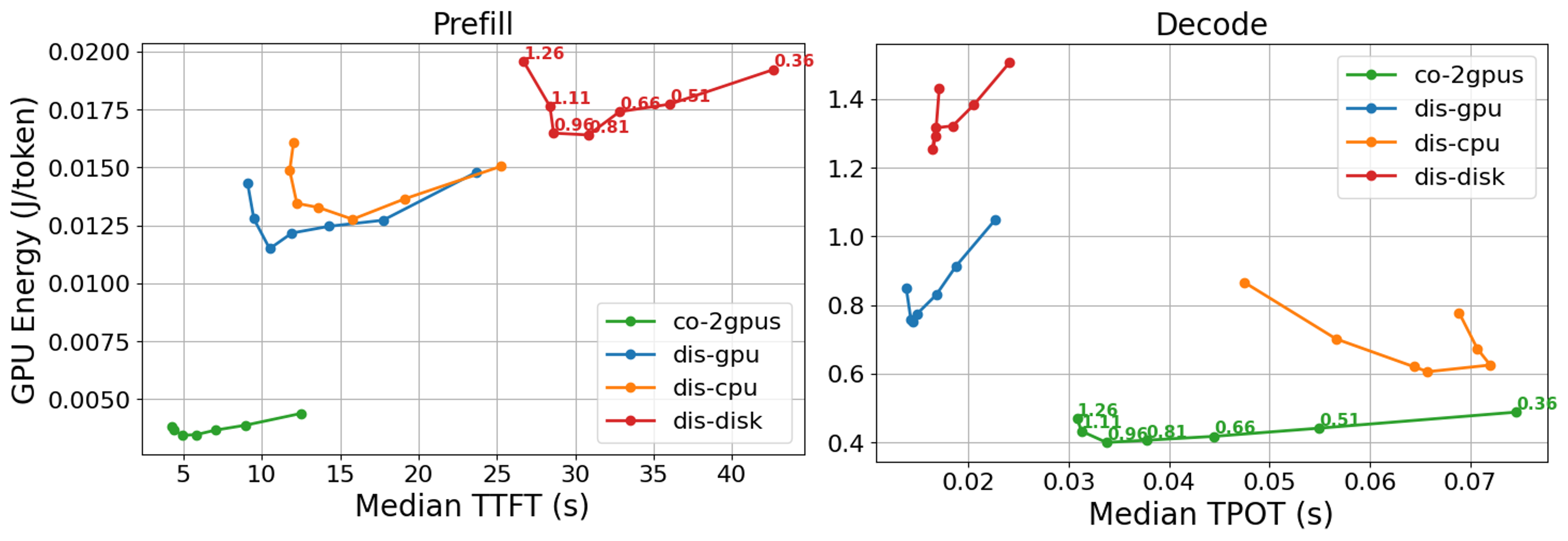}
    }
    \caption{TTFT-energy and TPOT-energy Pareto frontiers across different experimental setups. The inference workload includes a batch of 16 requests with input size 16,384 and output size 256.}
    \label{fig:exp2-latency}
\end{figure*}

We discuss the results of benchmarking experiment 2, which investigates the influence of GPU frequency scaling on performance and energy consumption among different experimental setups. The inference workload is fixed to be a batch of 16 requests with input size 16,384 and output size 256.

Figure \ref{fig:exp2-latency} shows the result, with the left sub-figure for prefill and the right for decode. The x-axis represents the latency performance, with Median TTFT for prefill and Median TPOT for decode. The y-axis indicates the total energy used for processing the inference workload. The labels associated with the lines represent the GPU frequency, measured in GHz. For all setups, we sample the frequency evenly from 0.36 GHz to 1.26 GHz and set to both GPUs, and measure the achieved latency and energy consumption for prefill and decode stages. This results in two latency-energy Pareto frontiers for each setup.

We notice that most latency-energy frontiers exhibit a U-curve. As GPU frequency increases, energy consumption decreases initially and increases ultimately. This is because of the marginal latency improvement from GPU frequency. Initially when GPU frequency is low, an increase in GPU frequency leads to a significant reduction in latency. During this period, the reduced latency counters the higher GPU power, and hence leading to decreasing energy consumption. As the GPU frequency approaches a certain threshold (e.g., 0.81 GHz), increasing GPU frequency leads to little improvement in latency. Instead, it increases GPU power and hence results in increasing energy consumption. These U-curve frontiers provide the opportunity to save energy consumption from frequency scaling. Through profiling those frontiers in the offline stage, we can potentially identify and set the GPU to the “sweet spot” frequency that can achieve the lowest GPU power consumption when latency is not constrained. Similarly, under the SLO-aware serving case where constraint is put on latency, we can search along the frontier to locate and set the GPU frequency that can satisfy the latency constraint at the lowest energy consumption.

We also notice that disaggregated serving setups consume substantially more energy than the colocated baselines. Even with independent frequency scaling, none of the disaggregated setups could outperform the colocated serving baselines in energy consumption. That leads to our second takeaway, independent frequency scaling does not guarantee energy savings. As discussed in the previous section, colocated serving exhibits better TTFT, and comparable TPOT when batch size is smaller. These improved latency performance ultimately leads to lower energy consumption in both prefill and decode stages. Although independent frequency scaling provides with more choices of TTFT and TPOT combinations, none of those combinations leads to a better energy consumption than that of colocated serving. However, though at the cost of higher energy consumption, we notice that \textit{dis-gpu} and \textit{dis-disk} provide improved TPOT than colocated serving, making them potential options for requests with tight TPOT requirements.

\section{Conclusion}
In this paper, we argue that existing prefill–decode disaggregated serving frameworks lack a comprehensive evaluation that compares performance and energy efficiency across different KV cache transfer paths. Moreover, although recent serving frameworks have supported optimization techniques such as KV cache reuse and frequency scaling, there remains a lack of systematic benchmarking to assess the performance and energy gains from these techniques under different conditions. To fill this research gap and provide a better understanding of the true performance and energy implications of prefill–decode disaggregation, we conduct a series of benchmarking studies that evaluate multiple experimental setups and optimization strategies. The major takeaways of our experiment are summarized below.

\begin{itemize}
    \item \textbf{Performance gains from prefill-decode disaggregation depends on the request load and KV transfer medium.} Compared to colocated serving, disaggregation doubles the request amount processed per GPU under equivalent GPU resources, resulting in higher overall TTFT when high-speed KV transfer interconnects are unavailable. However, at higher batch sizes, disaggregated serving outperforms colocated serving in TPOT, as colocated serving starts to suffer from cache eviction and recomputation due to limited GPU memory for KV cache.
    \item \textbf{Independent frequency optimization does not guarantee energy saving.} Across all KV transfer paths, disaggregated serving introduces longer prefill latency, which results in higher overall energy consumption compared to colocated serving. Although disaggregation enables independent frequency scaling to provide more choices of TTFT and TPOT combinations, none of the configurations ultimately achieve better energy consumption than the colocated baseline.
\end{itemize}

We hope those takeaways provide insights and guidance for future works that focus on prefill-decode disaggregated LLM inference serving and energy-aware LLM inference serving.

% \section*{Acknowledgment}

\bibliographystyle{IEEEtran}
\bibliography{references}

\end{document}